\documentstyle[12pt]{article}
\input{psfig}
\begin{document}
\vspace{3cm}
\title{\bf Entanglement and nonextensive statistics}
\date{}
\author{A. Vidiella-Barranco\thanks{Phone: +55 19 7885442; FAX:+55 19 7885427;
e-mail:vidiella@ifi.unicamp.br}\\
{\it Instituto de F\'\i sica ``Gleb Wataghin''}\\
{\it Universidade Estadual de Campinas}\\
{\it 13083-970   Campinas  SP  Brazil}.}
\maketitle
\begin{abstract}
It is presented a generalization of the von Neumann mutual information in the
context of Tsallis' nonextensive statistics. As an example, entanglement 
between two (two-level) quantum subsystems is discussed. Important changes 
occur in the generalized mutual information, which measures the degree of 
entanglement, depending on the entropic index $q$.
\end{abstract}

05.30.Ch; 03.67.-a; 89.70.+c 

\newpage

Entropy is undoubtely one of the most important quantities in physics.  
After more than one century of its introduction, it still generates
discussion around its nature and usefulness \cite{wer}. Of special 
importance it was the statistical interpretation of entropy given by Boltzmann, 
which allowed not only the development of classical statistical mechanics, 
but also the definition of its quantum mechanical counterpart, known as the von
Neumann entropy \cite{neum}. The von Neumann entropy associated to a 
quantum state of a system described by a density operator $\hat{\rho}$ 
is\footnote{Boltzmann's constant $k$ has been set equal to one here.}
\begin{equation}
S=-\mbox{Tr}\left[\hat{\rho}\ln\hat{\rho}\right] \ \ \ \ \ \ \mbox{Tr}\hat{\rho}=1.
\label{vnent}
\end{equation}
The von Neumann entropy does not depend on any of the system's observables,
being a function of the state itself. It is easy to see that if the 
above mentioned system is in a pure state 
$\hat{\rho}=|\Psi\rangle\langle\Psi|$, its entropy vanishes. Moreover, under
unitary evolution, entropy remains the same. However for statistical mixtures of
pure states we have that $S>0$, i.e., classical uncertainties increase the entropy
of the state. 

Recently Tsallis \cite{tsal} proposed a generalization of Boltzmann's entropy,
which in the quantum mechanical case reads
\begin{equation}
S_q=-\frac{1-\mbox{Tr}\left[\hat{\rho}^q\right]}{1-q}.
\end{equation}
The entropic index $q$ is a real parameter which is related to the (nonextensive)
properties of the relevant physical system.
In the limit of $q\rightarrow 1$, von Neumann's entropy is recovered. 
The Tsallis entropy has been succesfully applied to several interesting 
problems, involving nonextensive systems, which are normally untractable by 
means of Boltzmann's statistics. Amongst the problems treated within Tsallis'
formalism we may cite the L\'evy superdiffusion \cite{levy} and anomalous 
correlated diffusion \cite{corr}, turbulence in 2D pure electron plasma 
\cite{plas}, and in the analysis of the blackbody radiation \cite{blac}. There are 
more convenient 
values (or ranges of values) for the entropic index $q$, depending on the specific
system being treated. For instance, in the problem of thermalization in 
electron-phonon  systems, $q>1$ \cite{kopo}, while in the treatment of low 
dimensional dissipative systems, $q<1$ \cite{lyra}.

Entropy also plays a fundamental role in classical information theory \cite{shan}  
as well as in its quantum version. It may be considered as the average amount 
of information which is missing before observation. This may be quantitavely
expressed, for instance, through the Kullback-Leibler measure of information
\cite{kul}, sometimes called relative entropy. A generalization of this measure
and applications within Tsallis' statistics framework has been recently 
presented in the literature \cite{tkul,tkul1}, although in its classical version 
only. A discussion on channel capacities in nonextensive statistics may also be found
in the literature \cite{lands}.
Nevertheless there are normally not found in the literature discussions on the
implications of generalized statistics in purely quantum mechanical problems,
such as, for instance, on the measure of entanglement.

A quantity also used to compare distributions as well as quantum 
states is the mutual information or mutual entropy \cite{ohya}. The quantum
(von Neumann) mutual information $I$ relative to two subsystems ($A$ and $B$) 
may be written as
\begin{equation}
I=S_A+S_B-S_{AB},
\end{equation}
where $S_A(S_B)$ is the entropy relative to the subsystem $A(B)$, and $S_{AB}$ is
the entropy of the overall state, described by a density operator 
$\hat{\rho}_{AB}$. The reduced density operators relative to the subsystems,
$\hat{\rho}_{A}$ and $\hat{\rho}_{B}$ are obtained from $\hat{\rho}_{AB}$ through
the usual partial tracing operation, or
\begin{equation}
\hat{\rho}_{A}=\mbox{Tr}_{B}\hat{\rho}_{AB}, \ \ \ \ \ \ 
\hat{\rho}_{B}=\mbox{Tr}_{A}\hat{\rho}_{AB}.\label{partra}
\end{equation}
The von Neumann mutual information is then calculated using the quantum entropy
in (\ref{vnent}), i.e., $S_i=-\mbox{Tr}\left[\hat{\rho}_i\ln\hat{\rho}_i\right]$.
If the joint state $\hat{\rho}_{AB}$ is a pure 
state, then its von Neumann entropy $S_{AB}=0$, and according to the Araki-Lieb 
inequality \cite{araki} 
\begin{equation}
|S_A-S_B|\leq S_{AB}\leq S_A+S_B,
\end{equation}
we have that $S_A=S_B$, which means that the von Neumann mutual information 
is simply $I=2S_A$. In this pure state case, there are no classical uncertainties, 
and the correlations are purely quantum mechanical. Otherwise, if the joint state 
is a statistical mixture, $S_{AB}>0$, and there will be a mixing of classical and
quantum correlations. A convenient property of the von Neumann mutual 
information is that it is always positive definite. 

Quantum information theory has experienced a remarkable growth in the past years
\cite{qinf}, mainly motivated by potential applications in communication 
and computation. In particular, the adequate measure of quantum correlations and
entanglement \cite{vedral} is of central importance in this field. It would be 
therefore interesting to discuss a measure of correlations such as the mutual 
information, in a more general context. Here I present a straightforward
generalization of the von Neumann mutual information based on Tsallis
entropy ($S_q$), or
\begin{equation}
I_q=S_{qA}+S_{qB}-S_{qAB}.\label{gmi}
\end{equation}
This quantity would represent a generalization of the measure of correlations for a 
wider class of quantum systems (nonextensive). 

Now I am going to discuss an example involving a pair of two-state subsystems,
$A$ and $B$. The relevevant basis states will be denoted as $|0_\alpha\rangle$
and $|1_\alpha\rangle$ ($\alpha=A,B$), and $\langle i_\alpha|j_\alpha\rangle=\delta_{ij}$.
Let us assume that the overall system ($A$ plus $B$) is prepared 
in a state represented by the following state vector:
\begin{equation}
|\Psi\rangle_{AB}=p^{1/2}|0_A\rangle|1_B\rangle+(1-p)^{1/2}|0_B\rangle|1_A\rangle,
\end{equation}
with $0\le p \le 1$. If $p=0.5$ we have a maximally entangled state, and for $p=0$ (or
$p=1$) we have a disentangled (or product) state. We may even write a more general state,
which could include loss of coherence, as
\begin{eqnarray}
\hat{\rho}_{AB}&=&p|0_A\rangle|1_B\rangle\langle 1_B|\langle 0_A|
+(1-p)|1_A\rangle|0_B\rangle\langle 0_B|\langle 1_A|+\nonumber \\
&&\gamma^{1/2}[p(1-p)]^{1/2}\left(|0_A\rangle|1_B\rangle\langle 0_B|\langle 1_A|+ 
|1_A\rangle|0_B\rangle\langle 1_B|\langle 0_A|\right),
\end{eqnarray}
where the parameter $\gamma$ $(0\le \gamma \le 1)$ determines whether 
the state $\hat{\rho}_{AB}$ is a pure entangled state ($\gamma=1$), or a 
statistical mixture ($\gamma=0$). The partial tracing operation (\ref{partra}) 
produces the following density operators relative to the subsystems $A$ and $B$ 
\begin{equation}
\hat{\rho}_{A} = p|0_{A}\rangle\langle 0_A|+(1-p)|1_{A}\rangle\langle 1_A|,
\end{equation}
and
\begin{equation}
\hat{\rho}_{B} = (1-p)|0_{B}\rangle\langle 0_B|+p|1_{B}\rangle\langle 1_B|.
\end{equation}

The generalized mutual information for this particular system will then read
\begin{equation}
I_q=\frac{1}{q-1}\left[1+\eta_+^q+\eta_-^q-2\left(p^q+(1-p)^q\right)\right],
\end{equation}
where
\begin{equation}
\eta_\pm=\frac{1}{2}\left[1\pm\sqrt{1+4p(1-p)(1-\gamma)}\right].
\end{equation}

Now I analyze the behaviour of the generalized mutual information as a function
of the entropic parameter $q$. I shall remark that in the global pure state case 
($\gamma=1$), the mutual information exactly represents the degree of 
entanglement between the states belonging to the subsystems $A$ and $B$, due to 
the lack of ``classical noise''. In Fig. 1 we have a plot of the generalized
mutual information $I_q$ as a function of $q$ for different values of the parameter
$\gamma$, and with $p=0.5$, which corresponds to a
maximally entangled state in the case of $\hat{\rho}_{AB}$ being a pure state. 
For the range of values of $q$ here chosen $(0\le q \le 2)$ the generalized mutual 
information is positive-definite\footnote{Actually, $I_q$ is positive even for 
larger values of $q$.}. If $q=1$ we have the usual von Neumann mutual information.
For $\gamma=1$ (pure state), the mutual information decreases monotonically.  
It attains its maximum value, $I_q=2$, at $q=0$, and goes assimptotically to zero as
$q$ increases. However, an interesting behaviour is noticeable for states having
a small deviation from a pure state. For instance, if $\gamma=0.999$, although 
the von Neumann mutual information remains basically the same, ($I\approx 2\ln 2$),
the generalized
mutual information will substantially differ from that of a pure state 
($\gamma=1.0$) for not so large values of $q$ ($q<0.5$), as it is seen in Fig. 1. 
It starts increasing up to a maximum value at $q\approx 0.33$, then decreasing
again. In this case there will be in general two different values of $q$ giving the
same mutual information. 

We may also analyze the
behaviour of the generalized mutual information for a fixed $\gamma=1$ (pure
state) for different values of the weight $p$. If $p=1$ the state is a pure
disentangled one, having therefore mutual information $I_q$ equal to zero.
However, even for a very small ``entangled component'', for instance, if we take 
$p=0.999$, the generalized mutual information will assume nonzero values for a 
range of values of $q$. This is shown in Fig. 2. For $q=1$, $I_q\approx 0$, which 
means that according to von Neumann's mutual information, the state is viewed as
being completely disentangled. For other values of $q$, however, the generalized 
mutual information may be nonzero. This shows an extreme sensitivity of this 
measure of the degree of entanglement on the entropic index $q$, specially when
the quantum state is very close of being either a pure state or an entangled state. 
We conclude that entanglement may arise (or be enhanced), depending 
on the properties of a given physical system, such as extensivity, which is by its
turn quantified by the entropic index $q$.

The definition for the generalized mutual information here presented
($I_q$ in Eq. (\ref{gmi})) is not the only possible one. In fact, in 
\cite{tkul1} it is proposed, in its classical version, a slightly different form for a 
generalization of the mutual information. It would be therefore worth 
comparing our definition of mutual information in (\ref{gmi}) with the quantum 
mechanical counterpart of the one found in reference \cite{tkul1}, which may be written 
as
\begin{eqnarray}
I'_q&=&S_{qA}+S_{qB}-S_{qAB}+(1-q)S_{qA}S_{qB}=I_q+(1-q)S_{qA}S_{qB}\nonumber \\
&=&\frac{1}{q-1}\left[1+\eta_+^q+\eta_-^q-2\left(p^q+(1-p)^q\right)
-\left(1-p^q-(1-p)^q\right)^2\right].\label{gmial}
\end{eqnarray}
We note that the definition of the generalized mutual information given above 
contains an additional ``crossed term'' $(1-q)S_{qA}S_{qB}$ relatively to the 
definition we have already used ($I_q$). This might
result in some differences, which of course will depend on the values of the
parameters $p$ and $\gamma$. In the limit of $q\rightarrow 1$, von Neumann's mutual
information is recovered in either case. It would be convenient to perform a graphical 
comparison. For that I have plotted in Fig. 3 the generalized mutual information 
$I'_q$ from Eq.(\ref{gmial}) as a function of the entropic index $q$ having 
$p=0.5$, analogously to the situation depicted in Fig. 1. Despite of the 
differences we verify,
both definitions of the generalized mutual entropy, $I_q$ and $I'_q$ exhibit a very
similar qualitative behaviour. In particular, the important sensitivity discussed
above is present in both cases. The same is true in the case in which there is a 
small ``entangled component'' (analogous to the situation in Fig. 2). The 
corresponding plot for $I'_q$ is shown in Fig. 4, and again, its qualitative 
behaviour is about the same as the one in Fig. 2. This means that the discussion 
carried out above is also valid in the case of the alternative generalized mutual
information $I'_q$. 
In summary, the ``crossed term'' $(1-q)S_{qA}S_{qB}$ does not appreciably affect the 
(qualitative) behaviour of the generalized mutual information for those values of the 
parameters $p$ and $\gamma$ which are relevant, at least for what has been discussed
here.

I have presented a generalization of the quantum mechanical von Neumann's mutual 
information within Tsallis' nonextensive statistics. This observable-independent 
quantity, here denoted as $I_q$, is important for 
determining the degree of entanglement between different subsystems, for instance. 
I have found that depending on the value of the entropic index $q$ characteristic
of Tsallis statistics, the generalized mutual information, which measures quantum 
correlations may assume very different values from those obtained in the von 
Neumann case ($q=1$). The strong dependence of the mutual information on the 
entropic index may be of course associated to the extensivity properties of the 
relevant physical system, and entanglement arises for not so large 
values of $q$. I have also compared two different possible definitions of the
generalized mutual information, showing that the sensitivity relatively to the
entropic index $q$ is basically the same way in both cases.
This represents a first attempt of establishing a 
connection between an intrinsic property of physical systems (extensivity) and the 
measure of the degree of entanglement between different susbsystems.

\begin{figure}
\vspace{0.5cm}
\centerline{\psfig{figure=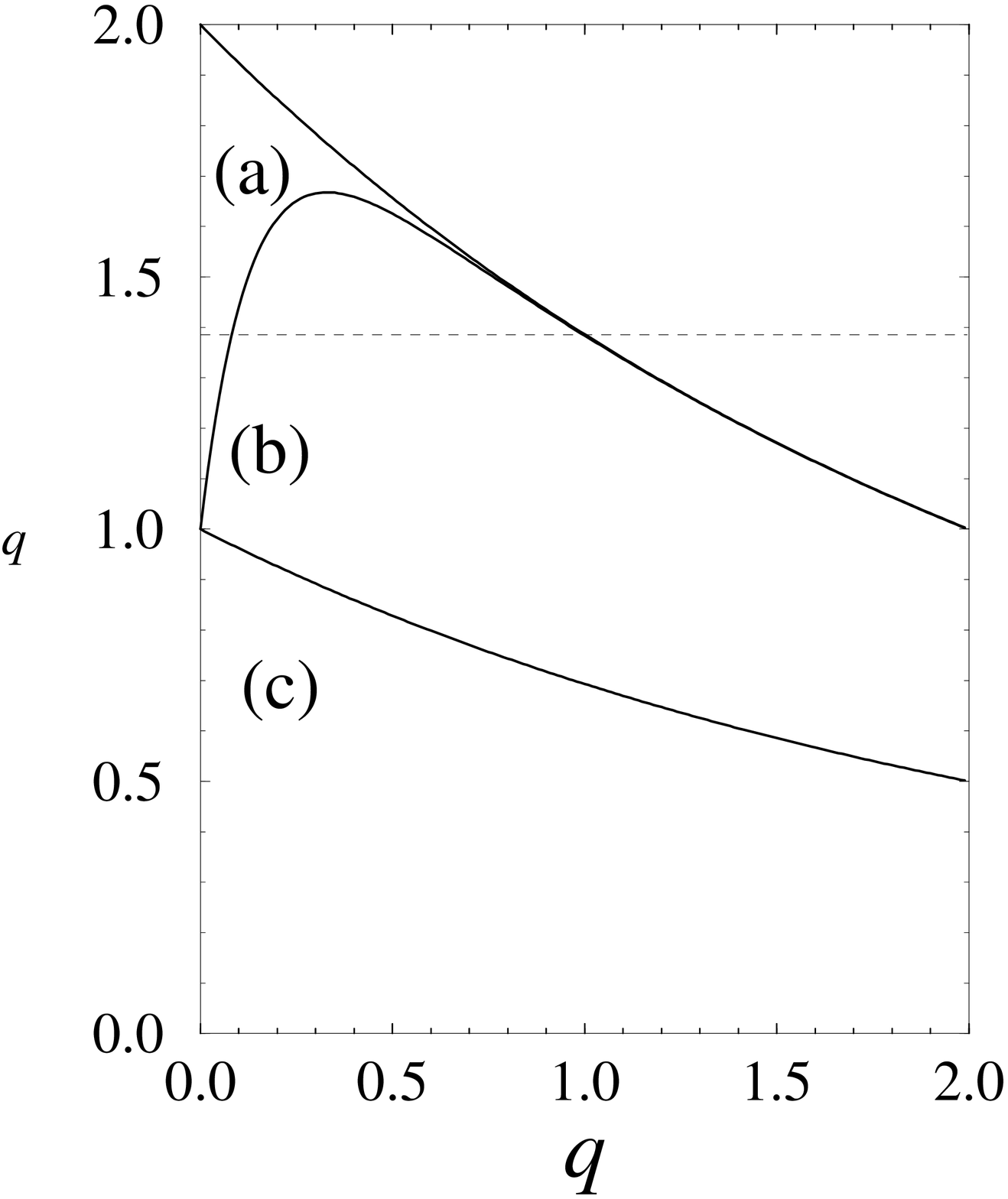,height=6.0cm,width=4.0cm}}
\vspace{1.5cm} 
\caption{Generalized mutual information $I_q$ as a function of the entropic index 
$q$ for different values of $\gamma$: (a) $\gamma=1$; (b) $\gamma=0.999$; (c) 
$\gamma=0$. In any case $p=0.5$. The dashed line indicates the von Neumann mutual 
information for the pure entangled state $I=2\ln 2$ ($\gamma=1$).}
\end{figure}

\vspace{1.5cm}

\begin{figure}
\vspace{0.5cm}
\centerline{\psfig{figure=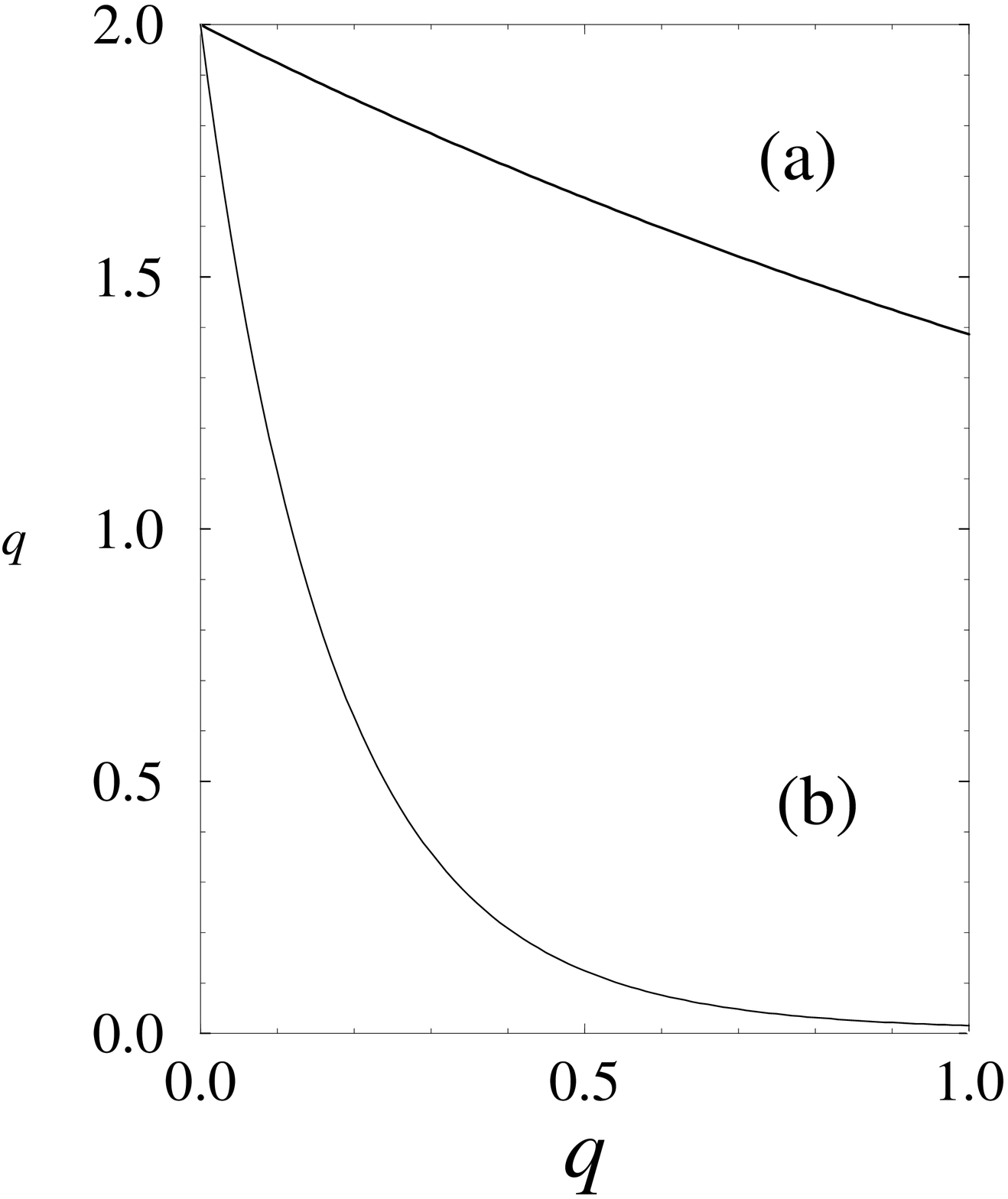,height=6.0cm,width=4.0cm}}
\vspace{1.5cm} 
\caption{Generalized mutual information $I_q$ as a function of the entropic index 
$q$ in the pure state case ($\gamma=1$) for (a) $p=0.5$; (b) $p=0.999$.}
\end{figure}

\newpage

\begin{figure}
\vspace{0.5cm}
\centerline{\psfig{figure=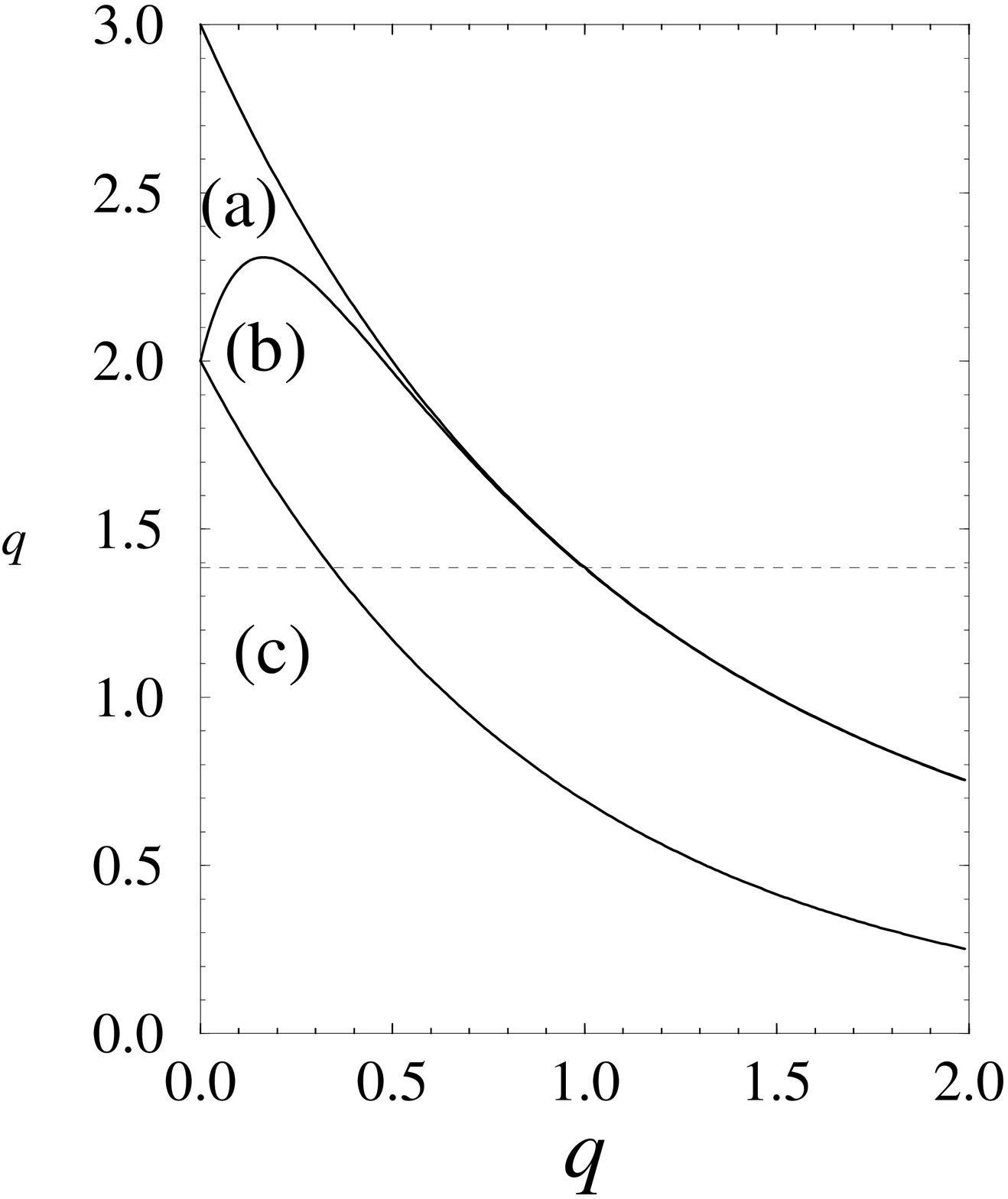,height=6.0cm,width=4.0cm}}
\vspace{1.5cm} 
\caption{Alternative generalized mutual information $I'_q$ as a function of the 
entropic index 
$q$ for different values of $\gamma$: (a) $\gamma=1$; (b) $\gamma=0.999$; (c) 
$\gamma=0$. In any case $p=0.5$. The dashed line indicates the von Neumann mutual 
information for the pure entangled state $I=2\ln 2$ ($\gamma=1$).}
\end{figure}

\vspace{1.5cm}

\begin{figure}
\vspace{0.5cm}
\centerline{\psfig{figure=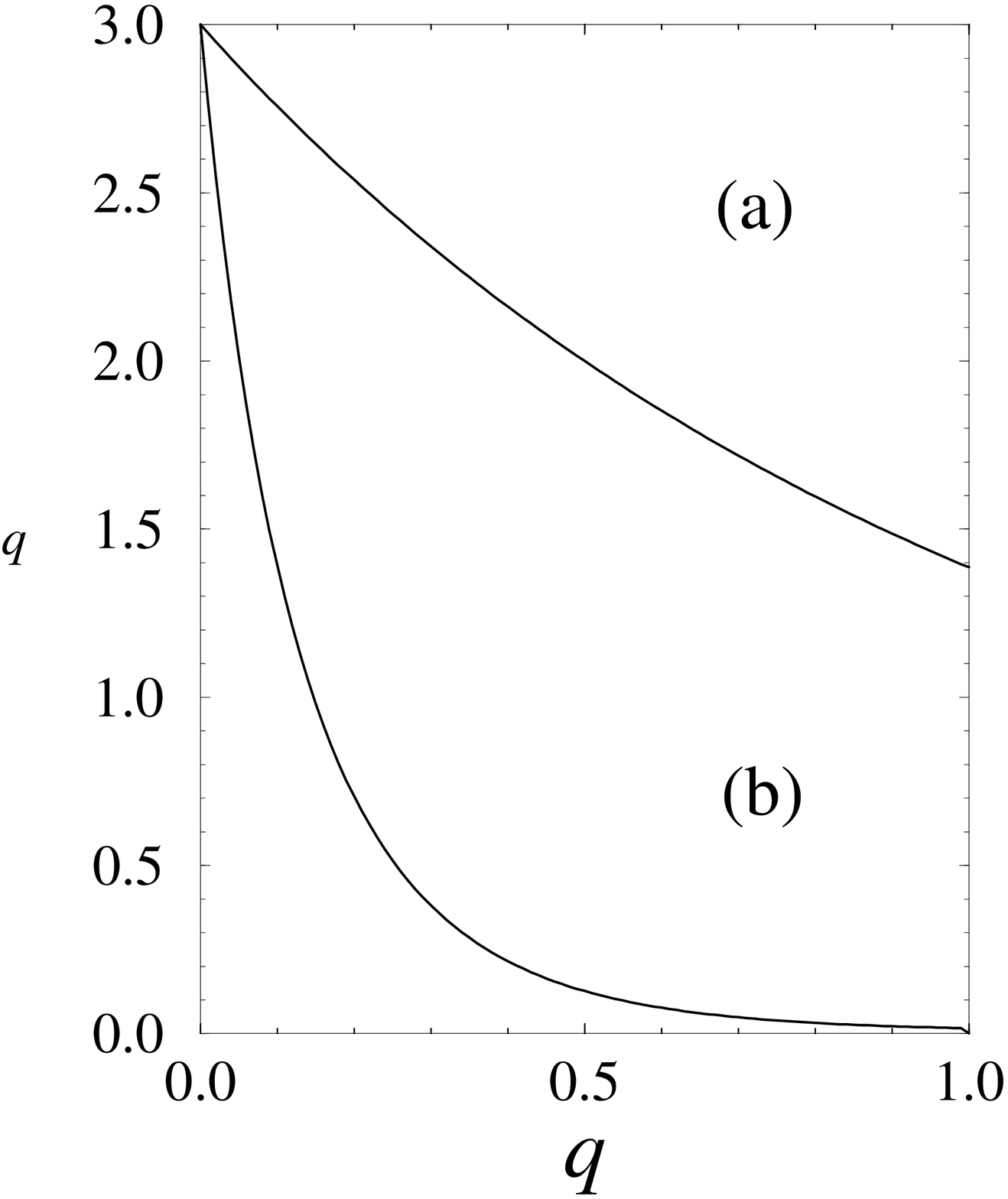,height=6.0cm,width=4.0cm}}
\vspace{1.5cm} 
\caption{Alternative generalized mutual information $I'_q$ as a function of the 
entropic index $q$ in the pure state case ($\gamma=1$) for (a) $p=0.5$; (b) 
$p=0.999$.}
\end{figure}

\end{document}